\documentclass[letter]{aa} 
\usepackage{graphicx}
\usepackage{txfonts}
\begin{document}
\newcommand{\bm}[1]{\mbox{\boldmath$#1$\unboldmath}}
\newcommand{\bmi}[1]{\mbox{\boldmath$\scriptstyle{#1}$\unboldmath}}
\renewcommand{\arcsec}{.\hspace{-0.9mm}'\!\hskip0.4pt'\hspace{-0.2mm}}
   \title{Evidence of convective rolls in a sunspot penumbra}


   \author{V. Zakharov
           \and
           J. Hirzberger
           \and
           T. L. Riethm\" uller
           \and
           S. K. Solanki
           \and
           P. Kobel
           }

   \offprints{V. Zakharov}

   \institute{Max-Planck-Institut f\" ur Sonnensystemforschung,
              37191 Katlenburg-Lindau, Germany\\
              \email{zakharov@mps.mpg.de}}

   \date{Received ; accepted }


  \abstract
   {}
   {We study the recently discovered twisting motion of bright penumbral
   filaments with the aim of constraining their geometry and the associated
   magnetic field.}
   {A large sunspot located $40\degr$ from disk center was observed at high
   resolution with the 1-m Swedish Solar Telescope. Inversions of
   multi-wavelength polarimetric data and speckle reconstructed time series of continuum
   images were used to determine proper motions, as well as the velocity and magnetic
   structure in penumbral filaments.}
   {The continuum movie reveals apparent lateral motions of bright and dark structures inside bright
   filaments oriented parallel to the limb, confirming recent Hinode
   results. In these filaments we measure upflows of $\approx 1.1~\mathrm{km/s}$ on their limbward side
   and weak downflows on their centerward side. The magnetic field in them is
   significantly weaker and more horizontal than in the adjacent dark filaments.}
   {The data indicate the presence of vigorous convective rolls in filaments with
   a nearly horizontal magnetic field. These are separated by filaments harbouring stronger,
   more vertical fields. Because of reduced gas pressure, we see deeper into
   the latter. When observed near the limb, the disk-centerward side of the horizontal-field
   filaments appear bright due to the \textit{hot wall} effect
   known from faculae. We estimate that the convective rolls transport most of the
   energy needed to explain the penumbral radiative flux.}


   \maketitle
%
%

\section{Introduction}
The discovery of twisting motions
of penumbral filaments seen in time series of filtergrams by Ichimoto et al.
(2007) ranks among the most striking recent discoveries in solar physics.
In sunspots observed at a heliocentric angle $\theta$ of $40-50\degr$ filaments
lying roughly parallel to the limb display a twisting motion
directed towards solar disk center. The nature and origin of this apparent
motion is still unclear and requires further study. An important step in
this direction is to determine how these twisting filaments fit into the
 complex magnetic structure of the penumbra. This is the main aim of the
 present letter, along with the confirmation of the Hinode-based results
 of Ichimoto et al. (2007) with the 1-m Swedish Solar Telescope, which allows
 higher spatial resolution images to be obtained.
%
%

\section{Observations and data reduction}
Using the 1-m Swedish Solar Telescope (SST), we observed a mature sunspot in active region NOAA
10904 on August 13, 2006 during a period of good to excellent seeing. The center of the field of view (FOV) was located
at $\mu=0.76$ (heliocentric angle $\theta=40.5\degr$). By means of a dichroic beamsplitter, the sunlight was split into red
($\lambda>500\mathrm{nm}$) and blue ($\lambda<500\mathrm{nm}$) beams.
In the blue channel, G-band ($\lambda_0=430.5\mathrm{nm}$ interference filter, FWHM=1.3~$\mathrm{nm}$), and
blue continuum ($\lambda_0=436.4\mathrm{nm}$, FWHM=1.1~$\mathrm{nm}$), images were
simultaneously collected on Kodak Megaplus 1.6 CCD cameras (1536$\times$1024 pixels, of 9~$\mathrm{\mu m}$
pixel-size). The exposure time was set to 13~$\mathrm{ms}$.
These images were binned into packages of 60 frames and processed using speckle image reconstruction techniques
(Weigelt 1977; Pehlemann \& von der L\"uhe 1989; de Boer 1996).
By applying this procedure, we obtained a two-hour long time sequence of near-diffraction-limited filtergrams at
a cadence of 19~s between individual frames.

In the red channel three temporally synchronized Sarnoff CAM1M100 cameras were employed, with exposure
times of 4.5~$\mathrm{ms}$. Two of them collected broad-band continuum frames near
$\lambda=$~630.25~$\mathrm{nm}$. The third camera registered spectro-polarimetric
data consisting of scans through the Fe~I ($\lambda=$~630.25~$\mathrm{nm}$) line at 6
wavelength positions
($\lambda-\lambda_0=[-150,-75,0,75,150,250]~\mathrm{m\AA}$) made with the Solar Optical Universal Polarimeter (SOUP)
filter. The last position in general samples the continuum. Two Liquid Crystal Variable Retarders (LCVRs) were used to modulate
the light beam. The complete scan across the iron line lasted about 123~$\mathrm{s}$.
After calibration procedures, the full Stokes vector at each point of the FOV was retrieved.
For this we used a telescope model and demodulation matrices of the optical setup measured with dedicated calibration
optics (Selbing 2005).

We determined the magnetic field vector and the line-of-sight (LOS)
velocity by inverting the measured Stokes profiles (composed of only 6 wavelength
points) using the HeLIx inversion code (Lagg et al.
2004). An atmosphere composed of two Milne-Eddington atmospheric components, one magnetic, the other one
representing unpolarized stray light, was employed. Extensive tests based on inverting synthetic
spectra computed in 3-D radiation
 MHD models have shown that 6 wavelength points are sufficient for a reliable inversion. \footnote{These tests
 were carried out by A.~Lagg et al. (unpublished) in the course of studying the design requirements of the Visible-light
 Imager \& Magnetograph for ESA's Solar Orbiter Mission.}
 The inversions provide the magnetic field vector: (field strength $|\vec{B}_{inv}|$,
 inclination $\gamma_{inv}$, azimuth $\phi_{inv}$) and the line-of-sight (LOS) velocity, $v_{\rm LOS}$, (which was calibrated to
be 0~$\mathrm{m/s}$ in the darkest part of the umbra).
We applied an algorithm based on minimization of the gradients of the azimuth over the FOV to correct for
the $180\degr$ ambiguity in $\phi_{inv}$. Additionally, $\vec{B}_{inv}$ was constrained to point away from the
sunspot umbra due to its positive polarity (confirmed by MDI magnetograms).
 If even after this algorithm for a given pixel $i$, $\Delta\phi_{inv}>90\degr$
($\Delta\phi_{inv}=|\phi_{inv,i}-\phi_{inv,i+1}|$), with respect to the $\phi_{inv}$ of 8 neighbouring pixels,
then we considered the results of the inversion of pixel $i$ to be
incorrect.
 The values in this pixel were then replaced by the median of the surrounding
 pixels. To analyse the geometry of the magnetic field better, a rotation matrix was
applied to $\vec{B}_{inv}$ to get the magnetic field vector $\vec{B}=\left(B,\gamma,\phi\right)$,
in local solar coordinates.

\section{Results}
Let us first consider the blue continuum
time sequence, which has the best spatial resolution of our
data. The most remarkable of the rapid intensity variations within penumbral
filaments are continuous lateral motions of dark and bright stripes across several bright filaments, giving a visual
impression of a twisting motion or a rotation around their axes.
To illustrate the temporal evolution of one such filament located in the
central penumbra, we plot in Fig.~1 a 228 $\mathrm{s}$ sequence of 4 blue continuum images
(1.5$\times$3.5~arcsec in size), with a cadence of $76$~$\mathrm{s}$. The filament
makes an angle of $\delta=-21\degr$ to the nearest limb (a negative number
implies that the outer part points towards disk center).
We filtered out the typical jitter of the
images, which comes from image distortions due to residual seeing effects, using a
subsonic filter that cuts out all motions faster than $4$~$\mathrm{km/s}$. In
Fig.~1  an elongated bright stripe moves from the left
side to the right of the filament, i.e. towards disk center. Another dark and bright
structures in this and other filaments move in the same direction. The average horizontal velocity of such motions is $\approx
1$~$\mathrm{km/s}$.
\begin{figure}
\includegraphics[width=8cm]{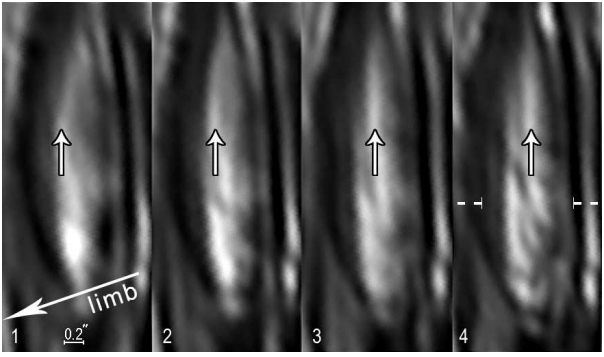}
\centering
\caption{Temporal evolution of an individual filament in blue continuum images.
The size of the FOV is 1.5$\times$3.5~arcsec and the cadence is $\Delta t=76$~$\mathrm{s}$.
White arrows mark the position of a bright structure that moves from left to right.
The solar limb is located to the left, the umbra to the bottom.}
\label{fig1}
\end{figure}
These lateral motions are always directed towards the disk-center side of the
filaments, regardless of their position in the penumbra (with respect to the umbra). These results agree with those obtained by Ichimoto et al. (2007).
Such motions can be identified over a large part of the penumbra. They are most prominent in filaments
oriented nearly parallel to the nearest limb ($\delta=0\degr$ or $180\degr$).
As the azimuth of the filaments to this direction
increases, the lateral motions become less visible. The effect vanishes at
angles $-140\degr\lesssim\delta\lesssim -40\degr$ and $40\degr\lesssim\delta\lesssim 140\degr$.

Next, we consider the results of inversions of the Stokes parameters.
In Fig.~2 we plot maps of normalized continuum intensity near
$\lambda=630.25\mathrm{nm}$, $I_{c}/I_{0}$ (where $I_{0}$ means the
continuum intensity averaged over the quiet Sun present in the same frame), $|\vec{B}|$,
$\gamma$, $\phi$, and $v_{LOS}$, of the same region as in Fig.~1. The magnetic
vector is given in solar coordinates. The dashed central contour
outlines the bright part of the filament ($I_{c}>1.0$) as is visible at the chosen viewing geometry.
The maps of $|\vec{B}|$, $\gamma$, $\phi$, and $v_{LOS}$
display an offset relative to the intensity map. Partly this is because
these are deduced from the $\lambda=630.25\mathrm{nm}$ line whose flanks are
formed typically $150~\mathrm{km}$ above $\tau_c=1$ in the penumbra, as we estimated
from contribution functions convolved with the filter profiles. To compensate
for this parallax, we need to shift the continuum contour by $\approx 100~\mathrm{km}$ to the left in
the 4 right panels in Fig.~2, which results in the solid contours. Even
after such a shift, a significant offset remains, which has a physical origin. Thus, the limb side
 of the filament shows blue-shifts, whereas its
center side displays weaker redshifts. The center-ward side of the filament has enhanced
magnetic field strength and greatly reduced zenith angle (i.e. a more vertical field),
 whereas an the limbward side the magnetic field is weaker and is almost horizontal.
 The azimuth at the limb side of the filaments is higher than at the disk-center
 side. Note that $\phi=0$ corresponds to a direction parallel to the $x$-axis and points
 towards the right, so that at the disk-center side the projection of the field is roughly
 parallel to the filament, while on the limbward side it points
 $\approx20\degr$ in the direction of the limb.

\begin{figure*}
\hspace{-17mm}
\includegraphics[angle=90, width=18.5cm]{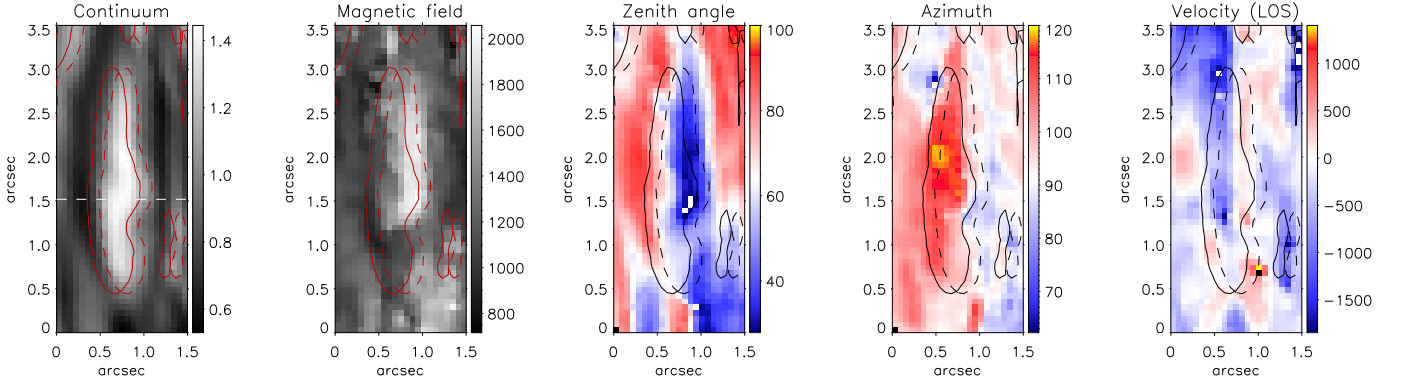}
\caption{\textit{From left to right:} continuum intensity near 630.25~$\mathrm{nm}$ normalized
to its value averaged over the nearest quiet Sun area,
magnetic field strength $|\vec{B}|$~[$\mathrm{G}$], zenith angle, $\gamma$~[$\degr$],
azimuth angle in the solar surface plane, $\phi$~[$\degr$], and the line-of-sight (LOS)
velocity, $v_{\rm LOS}$~[$\mathrm{m/s}$], respectively. Dashed contours localize the bright
penumbral structures ($I_{c}>1.0$). Solid contours have been shifted to roughly
compensate for the different heights to which continuum and the other physical
parameters refer. $\phi=0$ corresponds to a direction parallel to the $X$-axis
and points towards the right.}
\label{fig2}
\end{figure*}
\begin{figure}
\includegraphics[angle=0, height=18cm]{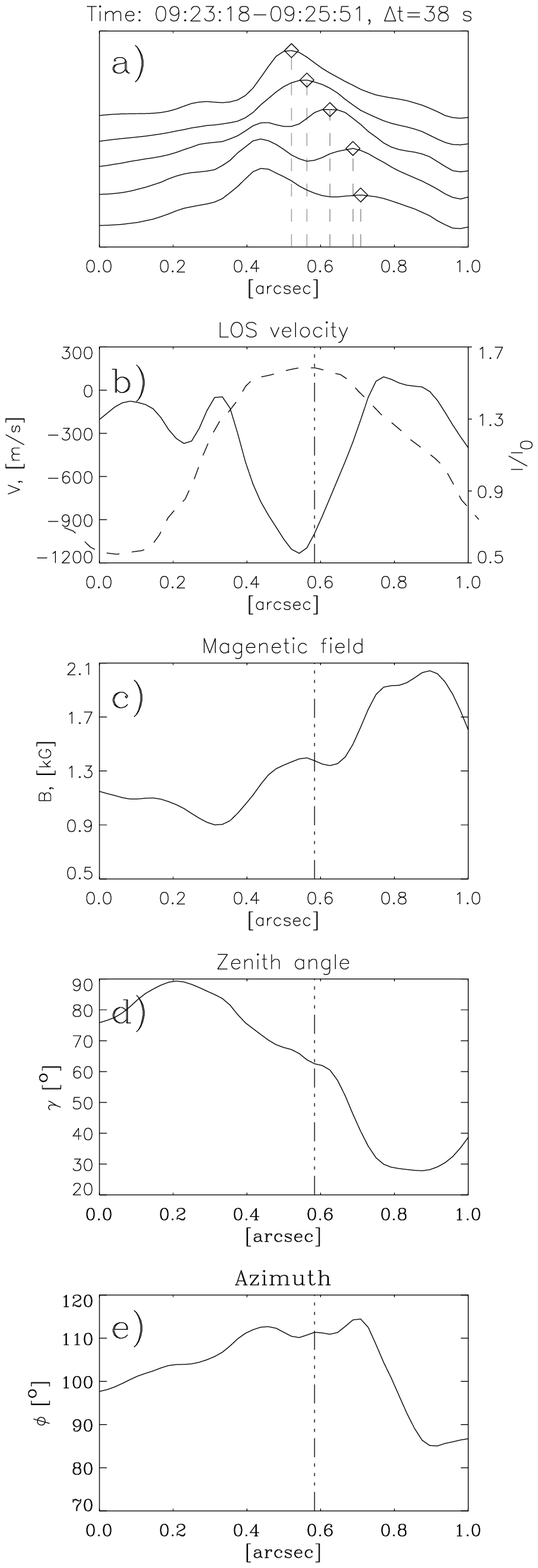}
\centering
\caption{\textit{From top to bottom:} a) horizontal cuts of blue continuum
intensity, b) LOS velocity (solid) and red continuum intensity (dashed), c) strength,
d) zenith angle, and e) azimuth of the magnetic field inside the filament. The solar limb is located to the
left. The continua profiles have been shiftet to the left by $0.1$~arcsec.
In panel a) intensity curves recorded later are offset downwards (cadence
$=38\mathrm{s}$). Diamonds mark the location of a
bright structure drifting across the filament. Panels b-e display the results of
data recorded between 09:16:50 - 09:18:30. The vertical
dashed-triple-dotted line corresponds to the location, averaged over the 09:16:50 - 09:18:30 time interval,
of red continuum intensity maximum of the bright filament.}
\label{fig3}
\end{figure}
Figure~3 displays a horizontal cut, made along the dashed line in Fig.~1
(frame~4) and in Fig.~2, of physical parameters. Figure~3a illustrates the
evolution of the blue continuum intensity along the cut. Lower curves were
recorded later at a cadence of 38~$\mathrm{s}$. 
Fig.~3b shows the variations of $v_{\rm LOS}$. The velocity profile shows an
 upflow (around $-1.13$~$\mathrm{km/s}$) on the limb side and
a weak downflow (around $+0.09$~$\mathrm{km/s}$) on the center side of the
filament.
This asymmetry between up- and downflows is also displayed by other
studied filaments. The velocity profiles presented by
Ichimoto et al. (2007) are also strongly
asymmetric, i.e. display a strong blue shift and a weak red shift.
This suggests that the asymmetry is an intrinsic property of the filaments,
although we cannot rule out that it is an artifact due to crosstalk from the Evershed
flow, a small component which may point in the direction of the LOS.

The continuum intensity (near$\lambda=630.25\mathrm{nm}$) too has been horizontally shifted by $100~\mathrm{km}$
to allow a parallax-free comparison with velocity
and magnetic field vector. The different spatial resolution and sensitivity to contrast
in the blue and red channels, as well as the fact that they were recorded at different
times, may be responsible for the rather distinct continuum profiles in Figs.~3a and~3b.
The center of the bright filament nearly coincides with the
highest blueshift.
These results are consistent with the findings of Ichimoto et al.
(2007).
Figures~3c,d,e display the components of the magnetic vector
$|\vec{B}|$, $\gamma$, and $\phi$, respectively.
The magnetic field strength reaches its maximum about $0.3$~arcsec towards
disk center from the axis of the bright filament. At this point $\vec{B}$ is
also most vertical ($\gamma=$28$\degr$) and
its azimuth is around 86$\degr$, which corresponds to an orientation of the
surface projection of the magnetic vector, $\vec{B}_{\rm{surf}}$,
roughly parallel to the filament. On the limb side of the bright filament and
in the dark region immediately bordering it, $|\vec{B}|$ is nearly a factor of 2
 lower and $\vec{B}$ becomes almost horizontal ($70\degr\lesssim\gamma\lesssim 90\degr$). Here the
 azimuth is skewed by $15\degr\lesssim\phi\lesssim 25\degr$ towards the limb relative to the axis of the filament.

\section{Discussion}
In Fig.~4 we present a sketch of the geometry of the penumbral fine
structure, which is consistent with the results of the present letter, as well as those of Ichimoto et al.
(2007), Borrero et al. (2008), and Borrero \& Solanki
(2008) based on Hinode data. The $\tau=1$ level is relevant for the line wings.
 In the sketched configuration stronger, more vertical fields (spines) are
associated with less dense gas, leading to a lowered $\tau=1$ level there.
Interspersed between them are filaments of nearly horizontal, weaker magnetic
field pointing into the page (interspines). Important is that the filaments carrying
the Evershed flow have a sizable horizontal field (Borrero \& Solanki 2008),
since the flowing gas is observed to be magnetized (Solanki et al. 1994;
Bellot Rubio et al. 2004). It is within these filaments that the transverse
motions are seen, which are similar to overturning convection. A nearly horizontal magnetic field means that
convection takes place in the form of convective rolls, as first proposed by Danielson (1961). Due to the
raised $\tau=1$ level in these filaments (consistent with the findings of
Rimmele 2008) the bright part of the filament, which is contoured in Fig.~2, is the
 penumbral counterpart of the \textit{hot wall}
found in faculae and pores (Spruit 1976; Keller et al. 2004; Carlsson et al. 2004; Cameron
et al. 2007). This geometry explains why only lateral motions directed
towards disk center are seen and why the observed upflow velocity is higher
than the downflow velocity, as can be seen by considering the directions of
the flow arrows relative to the line-of-sight in Fig.~4. From these hot walls
radiation flows into the neighbouring spines. The parallax effect, discussed
in Sect.~3, makes it obvious that the most inclined magnetic
field, which is located above the center of the filament, was detected at the
limbward border of the bright filament. The contrast between spine and
interspine is expected to depend on the relative temperature contrast, width
of the spine and the difference in evacuation between spines and interspines.
The values plotted in Figs.~2 and~3 are weighted averages along
the LOS (the relatively few measured wavelength points do not allow a more
detailed analysis). Thus, because a number of the slanted rays pass through
both the inclined and the horizontal fields, the difference between them
appears less clear in Figs.~2 and 3 than it may be in reality.
 Whether our measurements contradict the findings of Jur\v c\'ak \& Bellot Rubio
(2008), who report that \textit{bright penumbral filaments show the
more vertical fields and weaker flows}, cannot be judged, in part because of the different viewing geometry.
They have investigated the limbward part of the outer penumbra, whereas our
investigation addresses filaments oriented parallel to the limb, in the mid
penumbra.

The energy transport by the
convective rolls can be expressed as $F_{\rm cr}=\rho\cdot u\cdot v$, (neglecting the enthalpy) where
$\rho=2\times 10^{-7}\mathrm{g~cm^{-3}}$ is the mass density,
$u=3\times10^{12}~\mathrm{erg~g^{-1}}$ is the heat deposited by $1~\mathrm{g}$ of gas as it cools from $12000~\mathrm{K}$ to
$5000~\mathrm{K}$ (Schlichenmaier et al. 1999), and $v=1~\mathrm{km~s^{-1}}$ is the upflow velocity observed in
the convective rolls. Entering these numbers, we obtain
$F_{\rm cr}\approx 6\times10^{10}~\mathrm{erg~cm^{-2}~s^{-1}}$.
This is larger than the radiative flux emitted by the penumbra,
$F_{\rm pen}\approx 4.7\times10^{10}~\mathrm{erg~cm^{-2}~s^{-1}}=0.75\cdot F_{\sun}$.
However, we must keep in mind that upflows fill
at the most half of the surface area of the penumbra; e.g., the spines must
be heated radiatively. It is reasonable to assume them to emit the same flux
as the umbra in the absence of any lateral heat influx from the interspines,
$F_{\rm umb}=0.2\cdot F_{\sun}$. For a penumbra half covered by upflows and half
by gas at umbral temperatures, we obtain
the following relationship that must be fulfilled: $\frac{1}{2}\cdot\left(\rho\cdot u\cdot v+F_{umb}\right)=F_{\rm pen}$.
This gives a requirement on $\rho\cdot u\cdot v$ of $\approx 8
\times10^{10}~\mathrm{erg~cm^{-2}~s^{-1}}$, which is only a factor of
$1.3$ larger than the estimate obtained from the convective rolls. An
underestimation
of the partly unresolved velocities of the roll convection can easily account
for this factor. The observed roll convection transports far more energy flux
than interchange convection or the Evershed flow along flux tubes, as estimated by Schlichenmaier \& Solanki
(2003). We therefore propose that the observed convective rolls are the main form of
energy transport in the immediate subsurface layers of the penumbra, carrying
most of the energy required to maintain the radiative output of the penumbra.

The configuration of field and flows presented here combines aspects of both
(i) the uncombed penumbra model of Solanki \& Montavon (1993) and (ii) Schlichenmaier et al. (1998a,b) and the gappy penumbra model of Spruit \& Scharmer
(2006), Scharmer \& Spruit (2006), cf. Scharmer et al. (2008). The geometry of magnetoconvection in the umbral dots models by Sch\"ussler \& V\"ogler
(2006), cf. Riethm\"uller et al. (2008), displays qualitative similarities to our model. Observations by
Rimmele (2008) of a sunspot close to disk center reveal that the velocity pattern of
dark-cored filaments at lower layers is similar to what he sees in the
light bridges, i.e. the dark lanes are correlated with upflows and with downflows
to the sides. This also agrees with our results.
   \begin{figure}
   \includegraphics[width=8cm]{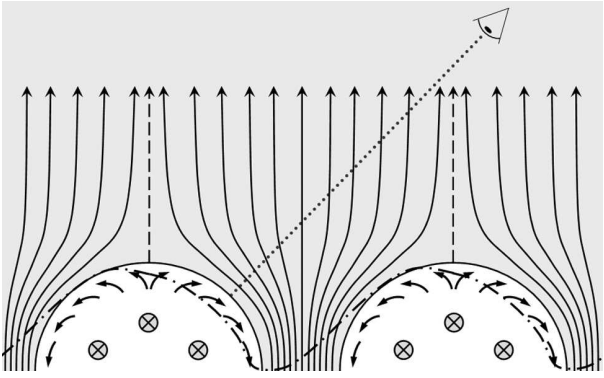}
   \centering
   \caption{Schematic illustration of the field and flow structure in penumbral
   filaments. Plotted is a vertical cut placed perpendicularly to the filaments
   (white semi circular shapes at the bottom of figure). The thick dashed-dotted
   line designates the $\tau=1$ level for the plotted view direction.
   Vertical arrows denote magnetic field lines and arrows inside the
   lightened filaments denote convective flows. Circled crosses represent
   magnetic fields inside the filaments oriented along their axes (which
   also correspond to the direction of the Evershed flow).}\label{fig4}
   \end{figure}

\begin{acknowledgements}
The Swedish 1-m Solar Telescope is operated on the island of La Palma by the
 Institute for Solar Physics of the Royal Swedish Academy of Sciences in the
  Spanish Observatorio del Roque de los Muchachos of the Instituto de Astrof{\'\i}sica
   de Canarias. We thank H.~Uthas, R.~Kever, and M.~van Noort for their kind support with the
   SST.
\end{acknowledgements}

%
%


\begin{thebibliography}{}
\bibitem[Bellot Rubio 2004]{bellot04} Bellot Rubio,~L.~R.,
  Balthasar,~H., \& Collados,~M. 2004, A\&A, 427, 319
\bibitem[de Boer]{boer96} de Boer,~C.~R. 1996, A\&AS, 120, 195
\bibitem[Borrero \& Solanki 2008]{borrero08b} Borrero,~J.~M., \& Solanki,~S.~K. 2008,
ApJ submitted
\bibitem[Borrero et al. 2008]{borrero08} Borrero,~J.~M., Lites,~B.~W., Solanki,~S.~K. 2008, A\&A, 481,
13
\bibitem[Cameron 2007]{cameron07} Cameron,~R., Sch\"ussler,~M., V\"ogler,~A., \&
Zakharov,~V. 2007, A\&A, 474, 261
\bibitem[Carlsson 2004]{carlsson04} Carlsson,~M., Stein,~R.~F., Nordlund,~\AA.,
Scharmer,~G.~B. 2004, ApJ, 610, 137
\bibitem[Danielson 1961]{danielson68} Danielson,~R.~E. 1961, ApJ, 134, 289
\bibitem[Ichimoto 2007]{ichimoto07} Ichimoto,~K., Suematsu,~Y., Tsuneta,~S. et al., 2007, Science, 318, 1597
\bibitem[Jur\v c\'ak \& Bellot Rubio 2008]{jurcak08} Jur\v c\'ak,~J, \& Bellot
Rubio,~L.~R. 2008, A\&A, 481, L17
\bibitem[Keller 2004]{keller04} Keller,~C.~U., Sch\"ussler,~M., V\"ogler,~A., \&
Zakharov,~V. 2004, ApJ, 607, 59
\bibitem[Lagg 2004]{lagg04} Lagg,~A., Woch,~J., Krupp,~N., \& Solanki,~S.~K.
2004, A\&A, 414, 1109
\bibitem[Pehlemann \& von der L\"uhe]{pehlemann89} Pehlemann,~E., von~der~L\"uhe,~O. 1989,
A\&A, 216, 337
\bibitem[Riethm\"uller et al. 2008]{rieth08} Riethm\"uller,~T.~L., Solanki,~S.~K., Lagg,~A. 2008, ApJ, 678, L157
\bibitem[Rimmele 2008]{rimmele08} Rimmele,~T. 2008, ApJ, 672, 684
\bibitem[Scharmer 2008]{scharmer08} Scharmer,~G.~B., Nordlund,~\AA.,
Heinemann,~T. 2008, ApJ, 677, 149
\bibitem[Scharmer 2006]{scharmer06} Scharmer,~G.~B., \& Spruit,~H.~C. 2006,
  A\&A, 460, 605
\bibitem[Schlichenmaier et al. 1998a]{schlichenmaier98a} Schlichenmaier,~R.,
  Jahn,~K., \& Schmidt,~H.~U. 1998a, ApJ, 493, 121
\bibitem[Schlichenmaier et al. 1998b]{schlichenmaier98b} Schlichenmaier,~R.,
  Jahn,~K., \& Schmidt,~H.~U. 1998b, A\&A, 337, 897
\bibitem[Schlichenmaier et al. 1999]{schlichenmaier99} Schlichenmaier,~R., Bruls,~J.~H.~M.~J., \&
Sch\"ussler,~M. 1999, A\&A, 349, 961
\bibitem[Schlichenmaier \& Solanki 2003]{schlichenmaier03} Schlichenmaier,~R., \& Solanki,~S.~K.
2003, A\&A, 411, 257
\bibitem[Sch\"ussler 2006]{schuessler06} Sch\"ussler,~M., \& V\"ogler,~A.
2006, ApJ, 641, L73
\bibitem[Selbing 2005]{selbing05} Selbing,~J. 2005, 'SST polarization model and polarimeter calibration', Master's thesis, Stockholm University
\bibitem[Solanki 1993]{solanki03} Solanki,~S.~K., \&
  Montavon,~C.~A.~P. 1993, A\&A, 275, 283
\bibitem[Solanki et al. 1994]{solanki04} Solanki,~S.~K., Montavon,~C.~A.~P.,
Livingston,~W. 1994, A\&A, 283, 221
\bibitem[Spruit 1976]{spruit76} Spruit,~H.~C. 1976, Sol. Phys., 50, 269
\bibitem[Spruit \& Scharmer 2006]{spruit06} Spruit,~H.~C., \& Scharmer,~G.~B. 2006,
  A\&A, 447, 343
\bibitem[Weigelt 1977]{weigelt77} Weigelt,~G.~P. 1977, OptCo, 21, 55
\end{thebibliography}
\end{document}